\documentclass[12pt]{article}
\usepackage{graphicx}

\newcommand\pubnumber{}
\newcommand\pubdate{\today}

\def\institute{Physikalisches Institut\\
Rheinische Friedrich-Wilhelms-Universit\"at Bonn, 53115 Bonn, GERMANY}

\def\Title#1{\begin{center} {\Large #1 } \end{center}}
\def\Author#1{\begin{center}{ \sc #1} \end{center}}
\def\Address#1{\begin{center}{ \it #1} \end{center}}

\newcommand\pubblock{\rightline{\begin{tabular}{l} \pubnumber\\
         \pubdate  \end{tabular}}}
\newenvironment{Abstract}{\begin{quotation}  }{\end{quotation}}
\newenvironment{Presented}{\begin{quotation} \begin{center} 
             PRESENTED AT\end{center}\bigskip 
      \begin{center}\begin{large}}{\end{large}\end{center} \end{quotation}}

\def\beq{\begin{equation}}
\def\eeq#1{\label{#1}\end{equation}}
\def\eeqn{\end{equation}}

\def\beqa{\begin{eqnarray}}
\def\eeqa#1{\label{#1}\end{eqnarray}}
\def\eeqan{\end{eqnarray}}

\let\bar=\overbar

\def\Dslash{\not{\hbox{\kern-4pt $D$}}}
\def\dslash{\not{\hbox{\kern-2pt $\del$}}}

\def\msb{{\bar{\ssstyle M \kern -1pt S}}}

\newcommand\pt{$p_T$}
\newcommand\mpt{\ensuremath{p_T}}
\newcommand*{\TeV}{\ifmmode {\mathrm{\ Te\kern -0.1em V}}\else
                   \textrm{Te\kern -0.1em V}\fi}%
\newcommand*{\GeV}{\ifmmode {\mathrm{\ Ge\kern -0.1em V}}\else
                   \textrm{Ge\kern -0.1em V}\fi}%

\begin{document}
\begin{titlepage}
\pubblock

\vfill
\Title{Boosted top: new algorithms and perspectives}
\vfill
\Author{ Julien Caudron,\\on behalf of the ATLAS and CMS Collaborations}
\Address{\institute}
\vfill
\begin{Abstract}
Studies of the boosted sector in top-quark physics have known a fast-growing development with the arrival of high-energy data at LHC. This short review summarizes the current status of the boosted top-tagging techniques in ATLAS and CMS and presents an overview of some of the most noticeable developments.
\end{Abstract}
\vfill
\begin{Presented}
$9^{th}$ International Workshop on Top Quark Physics\\
Olomouc, Czech Republic,  September 19--23, 2016
\end{Presented}
\vfill
\end{titlepage}
\def\thefootnote{\fnsymbol{footnote}}
\setcounter{footnote}{0}

\section{Introduction}

The top quark plays an important role in many beyond-the-standard-model scenarios.
When the top quark is produced with a very high momentum, its decay products are collimated, affecting the reconstruction efficiency of each individual decay product.
This specific phase space is called ``boosted regime'', and can be particularly interesting in some scenarios: for instance, when a new massive particle decays into a top quark pair, but also to observe effects of new physics in precision measurements. In addition, a boosted scenario study can complement a resolved scenario analysis by recovering events otherwise mis-reconstructed, or can provide additional advantages, such as a reduction of the combinatorial background due to less final state objects. 
Different techniques to identify high-\pt~top quarks based on substructure
analyses of large radius jets have been developed for Run-1 and Run-2
data by the ATLAS~\cite{atlas} and CMS~\cite{cms} collaborations together with the theory community.
This short review presents a selection of the current techniques and the recent developments that can possibly increase the performance of the top-tagging methods.

\section{Boosted selection and large-$R$ jets reconstruction}

The typical distance of two massless decay products of a heavy particle of mass $m$ is approximately $\Delta R = 2m/\mpt$.
The top quark decays almost exclusively into one $W$ boson and a bottom quark. The $W$ boson can decay hadronically into two jets or leptonically into one charged lepton and one neutrino.
In the latter case, in the boosted regime, the main concern is the charged lepton isolation, which is affected by the presence of the $b$-jet.
Isolation variables more robust to this effect are used, such as a cone isolation with a radius inversly proportional to the \pt~of the lepton, the tranverse lepton momentum with respect to the $b$-jet axis, or the distance between the $b$-jet and the lepton.
The majority of the efforts focuses on the hadronic decay case, where the most common method is to reconstruct a large-radius jet (large-$R$ jet) encompassing all the decay products. This large-$R$ jet is then considered as top-tagged (i.e., considered as the result of a top quark decay) based on the response of a specific algorithm, called ``top-tagger'', that studies the substructure properties of the jet.

\section{Current top-tagging algorithms}
The ATLAS and CMS collaborations have studied intensively several types of top-tagging algorithms~\cite{attp,cttp}.

The {\bf substructure variable taggers} are based on simple rectangular requirements on one or several substructure variables, such as 
the mass of the trimmed jet~\cite{krohn2010} or soft-dropped jet~\cite{softdrop},
the $k_t$ splitting scales~\cite{dij} obtained during the reclustering of the jet constituents with the $k_t$ algorithm,
the $N$-subjettiness ratios~\cite{thaler} obtained from the $N$-subjettiness variables $\tau_N$, the QJet volatility~\cite{qjet}, 
or the subjet $b$-tagging obtained with specific $b$-tagging algorithms more resilient to dense environment.

The {\bf HEPTopTagger} algorithm~\cite{htt} tests the compatibility of the hard structure of the large-$R$ jet
with the 3-prong pattern of the hadronic top quark decay.
Firstly, a mass drop criterion is used to decompose the large-$R$ jet into a collection
of subjets with mass lower than a given value. All possible triplets from this collection are then filtered to reduce contamination from underlying events
and pile-up and tested for compatibility with a hadronic top quark decay, based on the kinematic properties of the
reclustered three subjets and of the top candidate jet built from this procedure.
The large-$R$ jet is considered as tagged if there is at least one triplet satisfying this test. 
In the upgraded version~\cite{htt2}, used in the CMS collaboration, the procedure is iterated on different radius size, and the effective minimum radius can also be used as a discriminative variable.

The {\bf CMSTopTagger} algorithm~\cite{cmstt} is based on a decomposition procedure: for a given jet, the subjet pair of the last step of the clustering algorithm is used. The decomposition fails if the two subjets are too close. If one of the subjets is too soft, the soft subjet is rejected and the decomposition is applied to the other. This decomposition procedure is applied twice: once on the initial jet, and once on the possible two subjets created by the first iteration. The large-$R$ jet is considered as tagged if the number of final subjets is at least three and based on the kinematic properties of the subjets and the top candidate jet built from these subjets.

In the {\bf shower deconstruction} algorithm~\cite{sd},
likelihoods are calculated for the case the jet originates from the hadronic decay of a top quark (signal) and 
for the case the jet originates from a background process where hard gluons split into $q\bar{q}$ (background).
For signal and background, the likelihoods are calculated from first principles, including
the effect of the parton shower.
Subjets of the large-$R$ jet are identified with partons and a weight is calculated for each possible 
shower history that leads to the observed subjet configuration. 
This weight is proportional to the probability that the assumed initial particle generates the final 
configuration. 
A final discriminant variable $\chi$ is calculated as 
the ratio of the sum of the signal-hypothesis weights and the sum of the background-hypothesis weights.
The large-$R$ jet is considered as tagged if the $\chi$ value is higher than a given threshold.


\section{New developments}

The ATLAS collaboration has developed a {\bf variable-radius jet} algorithm~\cite{vrj} implementation, in which, during the jet reclustering algorithm, the beam closeness condition leading to the end of the iterative procedure is not fixed by a constant $R$ parameter, but by a variable $R_{eff}$ parameter, defined as $R_{eff} = \rho/p_T$ with $p_T$ being the transverse momentum of the current pseudo-jet, and $\rho$ being a parameter of the algorithm ($R_{eff}$ is also limited to the extremum values $R_{min}$ and $R_{max}$). The value of $\rho$ can be determined by computing the effective radius of a conventional jet in which the majority of the activity is contained. This study has shown that the substructure variable taggers using a jet clustered with this technique perform at least as well as the ones using the conventional clustering for a transverse momentum up to 1\TeV, and perform better for higher momentum.

The {\bf Heavy Object Tagger with Variable Radius} algorithm (HOTVR)~\cite{hotvr} has been developed independently by some CMS users. This algorithm is based on a modified jet clustering algorithm, using the variable-radius principle of the previous paragraph, and adding conditions to reject soft components during the clustering. Additionally, the identification of the subjets is done during the clustering, and can later be used in possible tagging criteria. This study has shown that, at generator level, using large-$R$ jets clustered with this algorithm provides competitive performance for the whole \pt~range.

The procedure to reduce the contamination from pile-up events in CMS has also been improved, using the {\bf pile-up per particle identification} algorithm (PUPPI)~\cite{puppi1,puppi2}.
This general technique can extend the particle-flow algorithm reconstruction and assign a weight to each particle-flow object. This weight is based on variables depending on the surrounding of the particle-flow object and is designed such that an object will have a weight closer to 1 if it is more likely associated to the leading vertex of the event and a weight closer to 0 if it is more likely associated to the pile-up contamination. The substructure variables of the large-$R$ jets produced after this cleaning procedure appear to have a better resolution, leading to a possible better discrimination when used in a top-tagger algorithm.

\begin{figure}[htb!]
   \begin{centering}
      \includegraphics[width=0.4\textwidth]{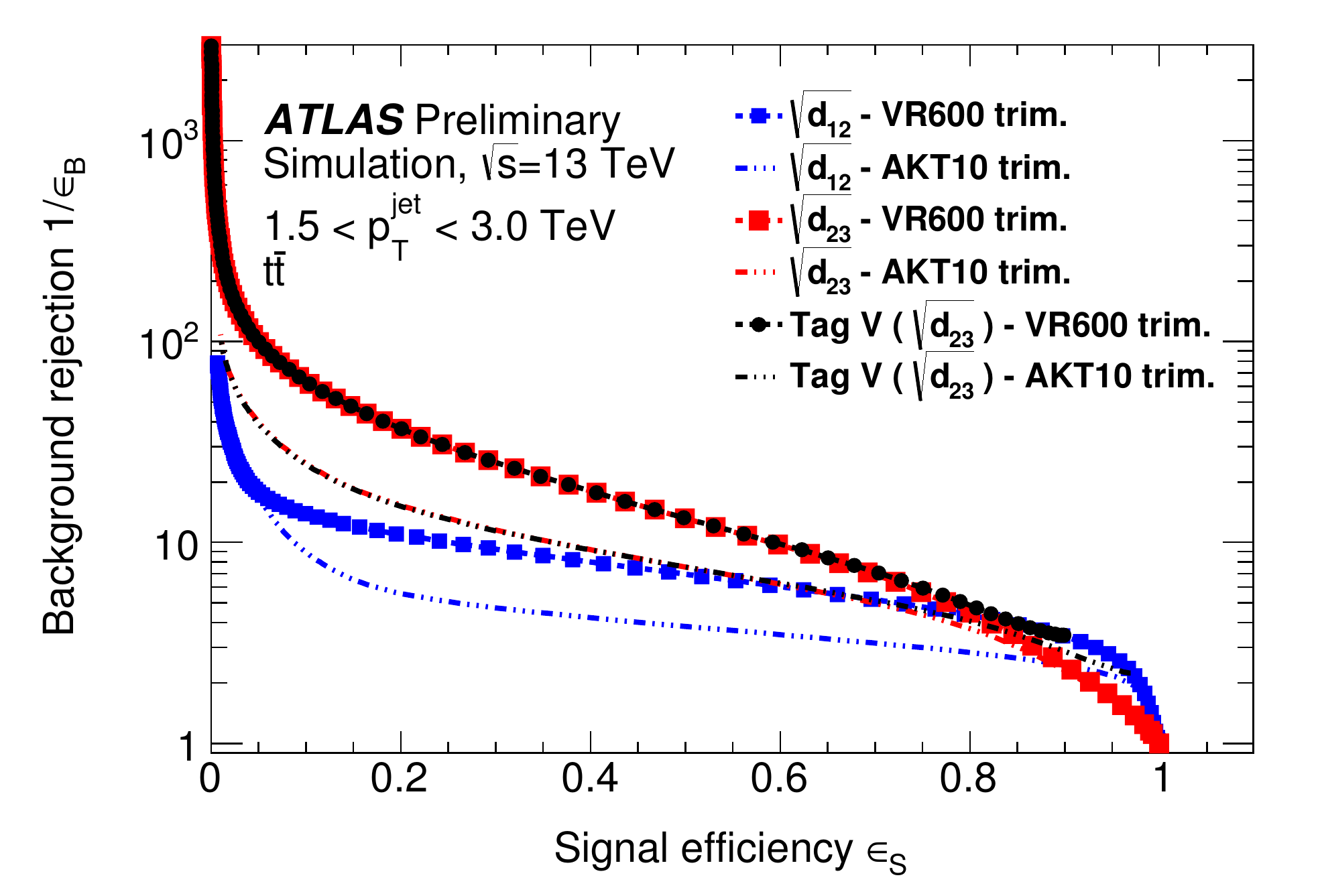}
      \includegraphics[width=0.4\textwidth]{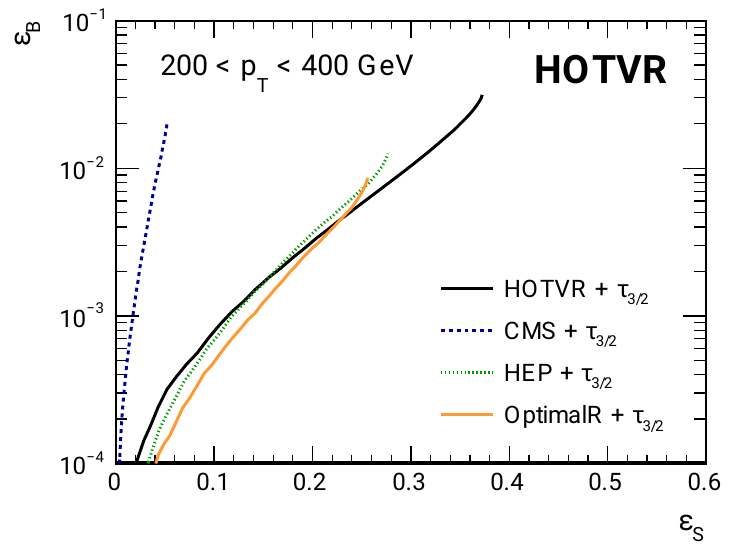}\\
      \includegraphics[width=0.4\textwidth]{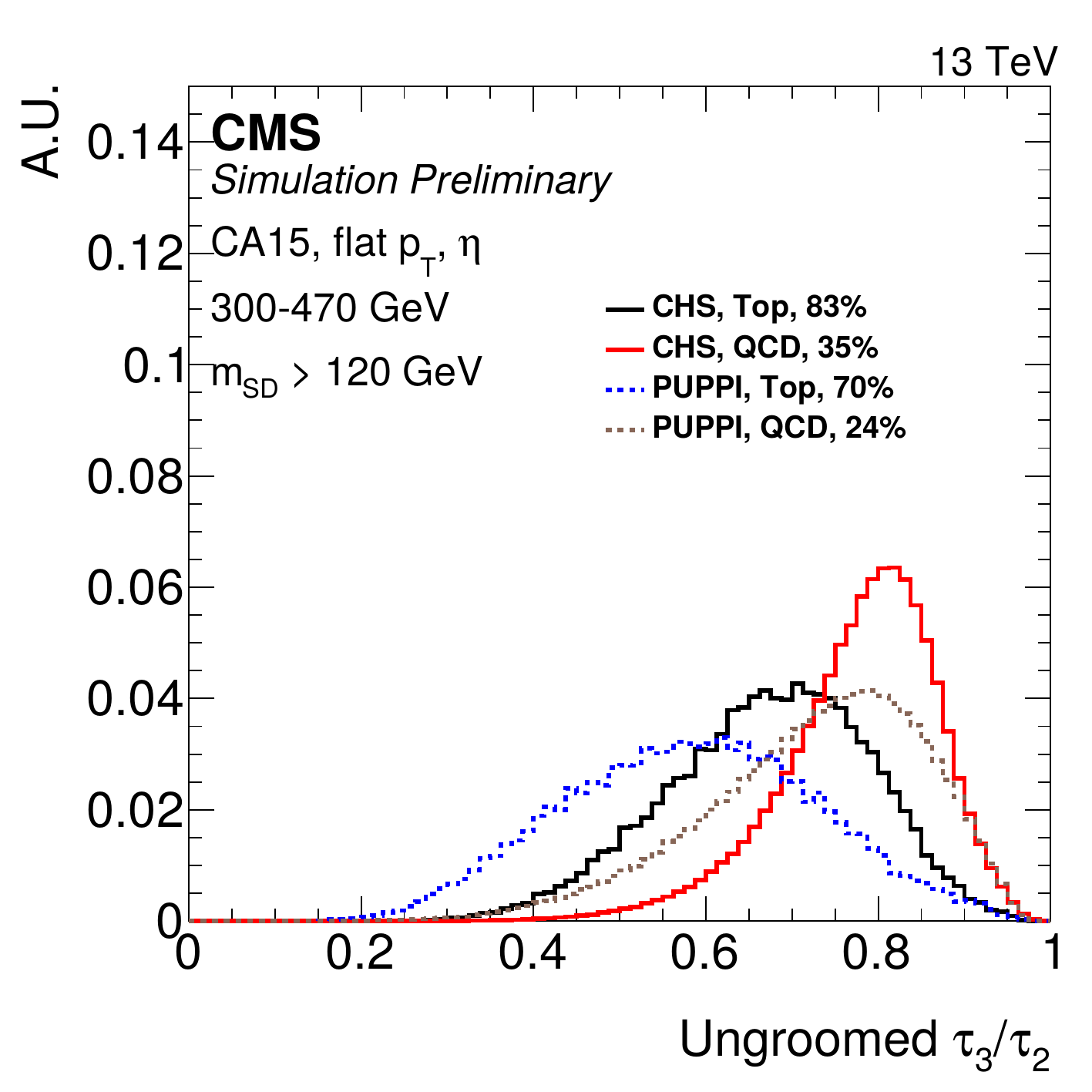}
      \includegraphics[width=0.4\textwidth]{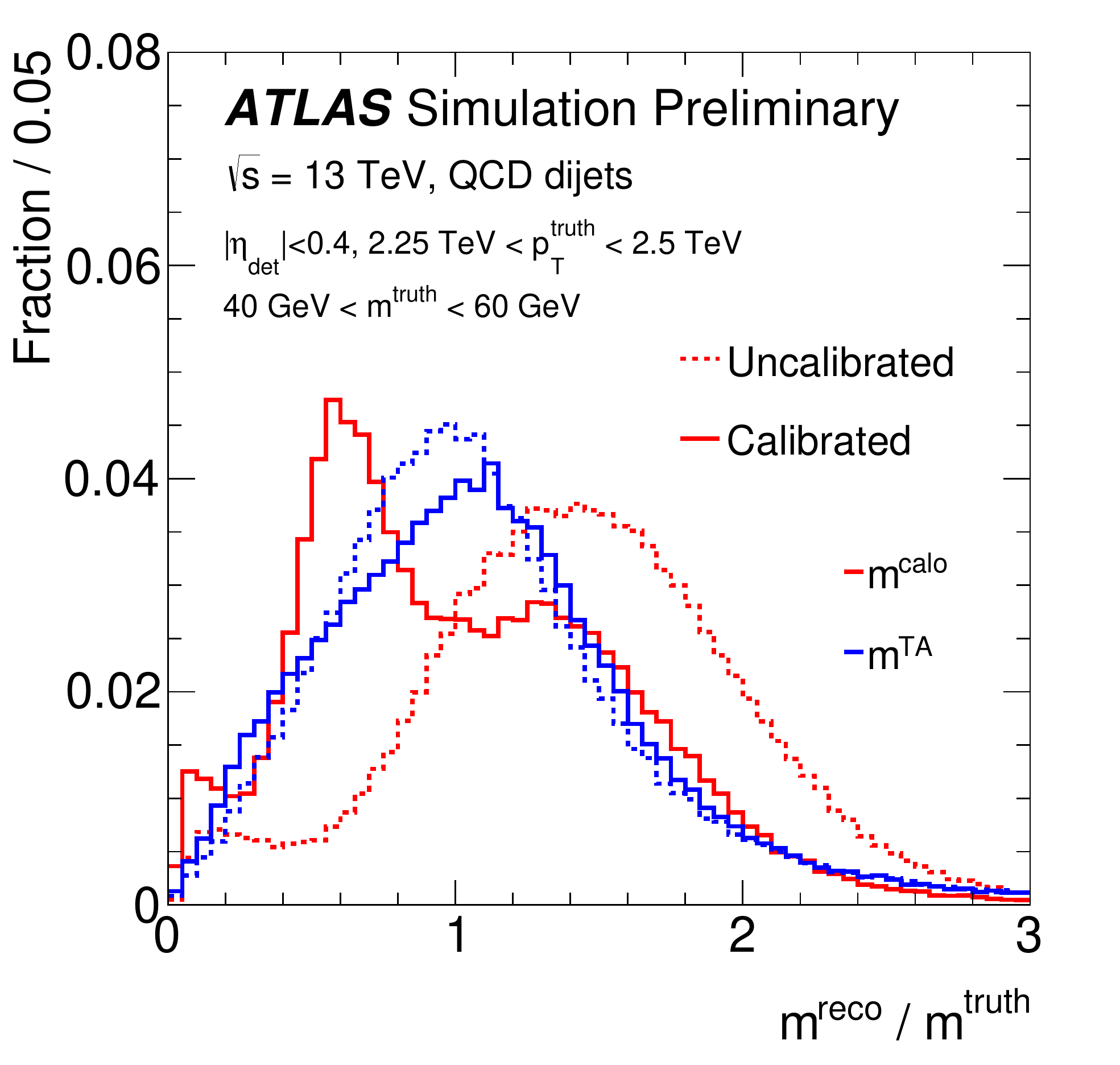}\\
      \caption{
        Illustrations, for each technique, of the possible gain: (a) better performance for high momentum for the variable-radius jet algorithm~\cite{vrj}, (b) competitive performance for the heavy object tagger with variable radius~\cite{hotvr}, (c) a better resolution of the discriminating variables for the pile-up per particle identification algorithm~\cite{puppi2}, (d) and a good behavior of the calibration for low mass high momentum jets with the track-assisted mass reconstruction~\cite{mta}. See references for more details.
      }
      \label{fig:new}
   \end{centering}
\end{figure}

Improvements have also been achieved in ATLAS in the jet mass reconstruction, using the {\bf track-assisted mass reconstruction} algorithm~\cite{mta}.
The usual procedure exploits the deposit in the calorimeters, but the calibration procedure starts to fail for high momentum jets and low mass. This is due to the limited granularity of the calorimeter. The track-assisted mass, which is the mass reconstructed only from the tracks associated to the jet and corrected to account for the neutral particles, does not suffer from this limitation. In the context of top-tagging, this procedure does not improve the mass resolution of large-$R$ top jet (due to the large top mass), but it can possibly improve the performance of a top-tagging algorithm if this algorithm relies on the subjet masses and if this procedure is applied to these subjets.

\section{Conclusion}

In this short review, four current techniques for identifying a large-radius jet as the result of a top quark decay (top-tagging) have been presented (substructure variable taggers, HEPTopTagger, CMSTopTagger and shower deconstruction). The performance of those techniques depends of the top quark transverse momentum range and of the top-jet identification efficiency and mis-identification rate relevant for a given analysis.
This short review has also presented four recent developments in top-tagging, jet reconstruction or event reconstruction techniques (variable-radius jet, heavy object tagger with variable radius, pile-up per particle id and track-assisted mass reconstruction) that can possibly improve the performance of the top-tagging. Illustrations of these possible improvements are given in Figure~\ref{fig:new}.

\end{document}